\documentclass[aps,prl,twocolumn,amsmath,amssymb,superscriptaddress,showkeys,showpacs]{revtex4}
\usepackage[dvips]{graphicx,color}
\usepackage[dvips]{hyperref}
\hypersetup{
  pdftitle={Determination of the $\eta^{\prime}$-proton scattering length in free space}
  pdfauthor={Eryk Czerwi{\'n}ski, Pawe{\l} Moskal, Micha{\l} Silarski, Steven Bass et al.},
  pdfkeywords={total width, eta-prime, missing mass},
  unicode=false,
  colorlinks=false,
}
\usepackage[english]{babel}
\usepackage[sort&compress]{natbib}

\begin{document}
\title{Determination of the $\eta^{\prime}$-proton scattering length in free space}
\author{E.~Czerwi\'nski}\email[Electronic address: ]{eryk.czerwinski@uj.edu.pl}
\affiliation{Institute of Physics, Jagiellonian University, PL-30-059 Cracow, Poland}
\author{P.~Moskal} 
\affiliation{Institute of Physics, Jagiellonian University, PL-30-059 Cracow, Poland}
\author{M.~Silarski}
\affiliation{Institute of Physics, Jagiellonian University, PL-30-059 Cracow, Poland}
\author{S. D.~Bass}
\affiliation{Stefan Meyer Institute for Subatomic Physics,
Austrian Academy of Sciences, Boltzmanngasse 3, A 1090 Vienna, Austria}
\author{D.~Grzonka}
\affiliation{Institute for Nuclear Physics and J{\"u}lich Center for Hadron Physics,\\
                Research Center J{\"u}lich, D-52425 J{\"u}lich, Germany}
\author{B.~Kamys}
\affiliation{Institute of Physics, Jagiellonian University, PL-30-059 Cracow, Poland}
\author{A.~Khoukaz}
\affiliation{IKP, Westf\"alische Wilhelms-Universit\"at, D-48149 M\"unster, Germany}
\author{J.~Klaja}
\affiliation{Institute of Physics, Jagiellonian University, PL-30-059 Cracow, Poland}
\author{W.~Krzemie{\'n}}
\affiliation{Institute of Physics, Jagiellonian University, PL-30-059 Cracow, Poland}
\author{W.~Oelert}
\affiliation{Johannes Gutenberg-Universit{\"a}t Mainz, 550099 Mainz, Germany}
\author{J.~Ritman}
\affiliation{Institute for Nuclear Physics and J{\"u}lich Center for Hadron Physics,\\
                Research Center J{\"u}lich, D-52425 J{\"u}lich, Germany}
\author{T.~Sefzick}
\affiliation{Institute for Nuclear Physics and J{\"u}lich Center for Hadron Physics,\\
                Research Center J{\"u}lich, D-52425 J{\"u}lich, Germany}
\author{J.~Smyrski}
\affiliation{Institute of Physics, Jagiellonian University, PL-30-059 Cracow, Poland}
\author{A.~T\"aschner}
\affiliation{IKP, Westf\"alische Wilhelms-Universit\"at, D-48149 M\"unster, Germany}
\author{M.~Wolke}
\affiliation{Department of Physics and Astronomy, Uppsala University, SE-751 20 Uppsala, Sweden}
\author{M.~Zieli\'nski}
\affiliation{Institute of Physics, Jagiellonian University, PL-30-059 Cracow, Poland}
\date{\today}
\begin{abstract}
Taking advantage of both the high mass resolution of the COSY--11 detector and the high energy resolution of the low-emittance 
proton-beam of the Cooler Synchrotron COSY we determine 
the excitation function
for the $pp\to pp\eta^{\prime}$ reaction close-to-threshold.
Combining these data with previous results we extract
the scattering length for the $\eta^{\prime}$-proton potential 
in free space to be
$\mathrm{Re}(a_{p\eta'}) =  0~\pm~0.43~\mathrm{fm}$ and
$\mathrm{Im}(a_{p\eta'}) = 0.37^{~+0.40}_{~-0.16}~\mathrm{fm}$.
\end{abstract}
\maketitle
In this letter we report the determination of the scattering 
length for the interaction of the $\eta^{\prime}$ meson 
with the proton based on the shape of the excitation function 
for the $pp\to pp\eta^{\prime}$ reaction measured close to the 
kinematic threshold.
Using the high mass resolution of the updated COSY--11 
detector~\cite{Brauksiepe,Moskal22} and the low-emittance 
proton-beam of the Cooler Synchrotron COSY~\cite{Maier} 
the excitation function 
was determined 
down to excess energy $Q = 0.76$~MeV above threshold, 
with the precision 
$\Delta{Q}=0.1$~MeV improved by more than a factor of five 
with respect to previous measurements.
The improved 
resolution enabled quantitative extraction of 
the $\eta^{\prime}$ proton scattering length in free space.

The scattering lengths describing interaction potentials between mesons and nucleons are of fundamental importance in hadron physics.
However, they are not well established especially
for
those flavor neutral mesons that are characterized by very short life
times
making investigations of the meson-nucleon potential 
in the standard way via scattering experiments impossible.
So far,
based
on the shift and width of the ground state of pionic hydrogen 
atoms~\cite{pi0-Sigg},
only the scattering length of the $\pi^0$-nucleon potential 
is accurately determined
with a precision of about 0.001~fm.
The scattering length for the $\eta$-nucleon potential is determined
more than two orders of magnitude less precisely, with phenomenological values quoted for the real part between $\sim$0.2~fm and 
$\sim$1~fm depending on the analysis method~\cite{arndt05}.
Until now the $\eta^{\prime}$-nucleon scattering length had 
been estimated only qualitatively~\cite{Moskal9}.

Measurements of the $\eta$- and $\eta'$- nucleon and nucleus 
systems are sensitive to dynamical chiral and axial U(1) 
symmetry breaking in low energy QCD.
While pions and kaons are would-be Goldstone bosons associated 
with chiral symmetry, the isosinglet $\eta$ and $\eta'$ mesons 
are too massive by about 300-400 MeV for them to be pure 
Goldstone states.
They receive extra mass from non-perturbative gluon dynamics 
associated with the QCD axial anomaly.
This OZI violation is also expected 
to influence the $\eta'$-nucleon interaction \cite{bass99}.
Without the gluonic mass contribution the $\eta'$ 
would be a strange quark state after $\eta$-$\eta'$ mixing
(and the $\eta$ would be 
 a light-quark state degenerate with the pion), 
mirroring the situation with isoscalar $\phi$ and $\omega$ 
vector mesons.  
To the extent that coupling to nucleons and nuclear matter is
induced by light-quark components in the meson,
any observed scattering length and mass shift in medium 
is induced by the QCD axial anomaly that generates part of 
the $\eta'$ mass \cite{cracow13}.

\begin{figure}[h]
  \begin{center}
  \vspace{-5mm}
    \includegraphics[width=0.47\textwidth]{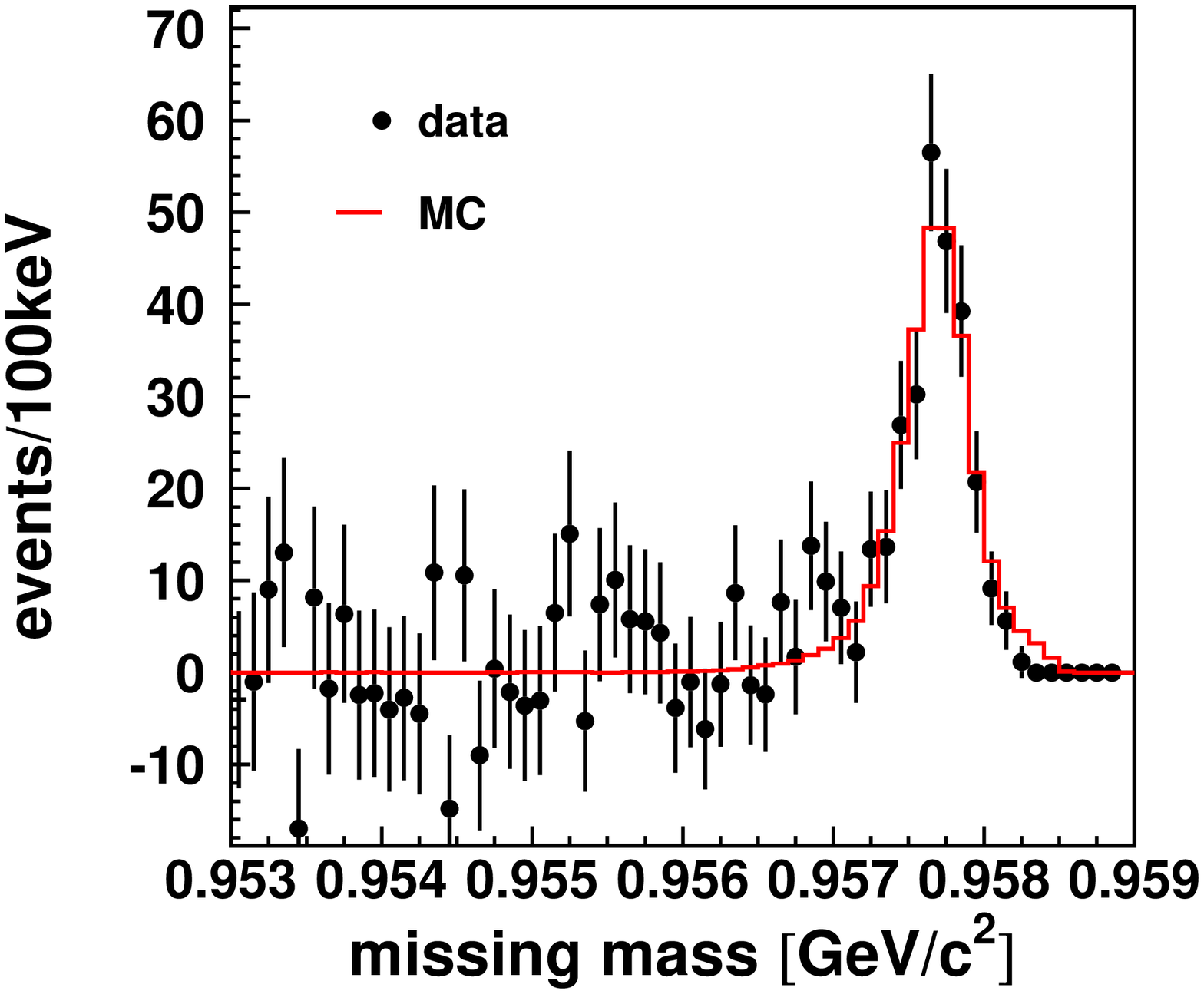}
\\
  \vspace{-7mm}
    \includegraphics[width=0.47\textwidth]{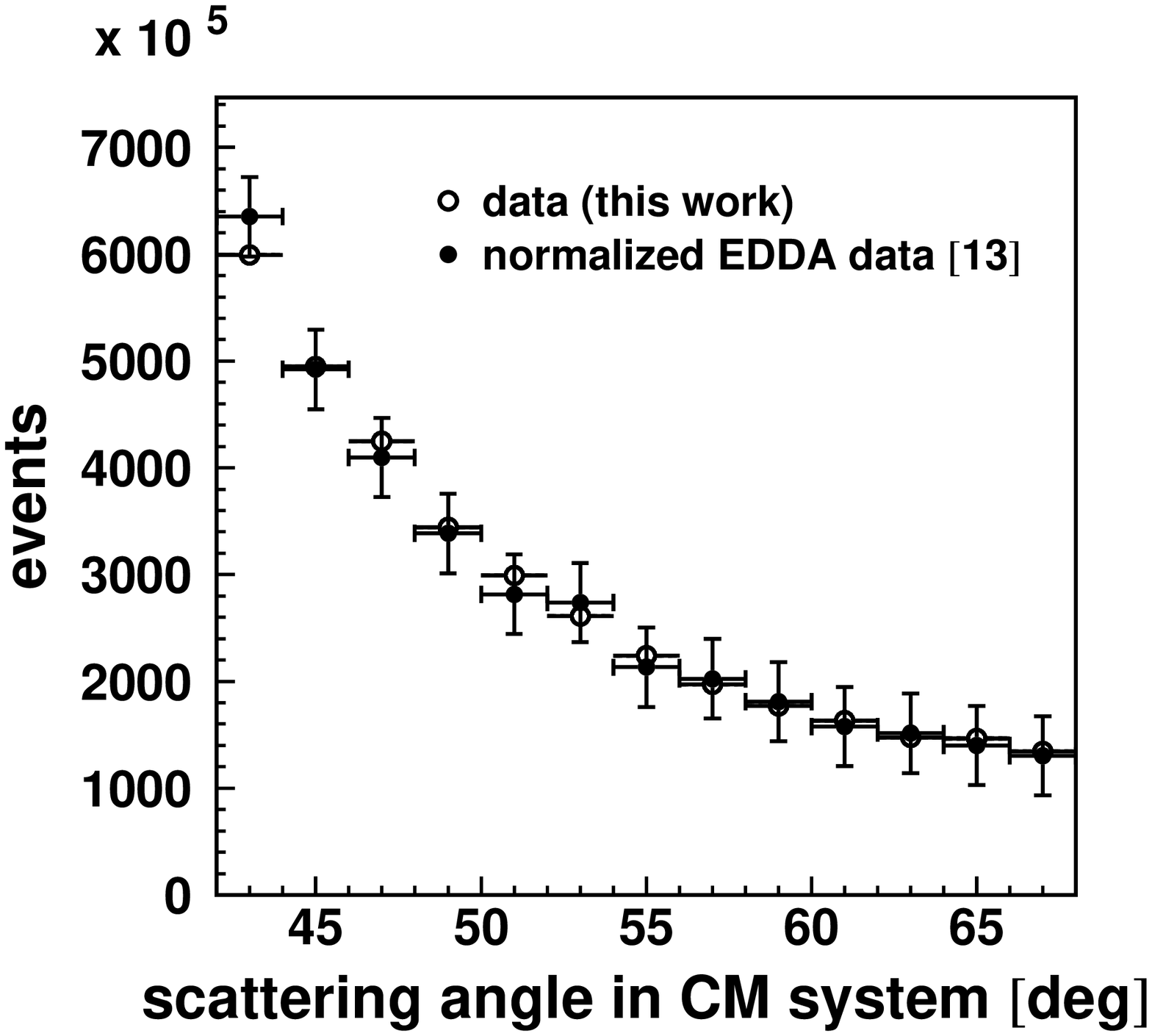}
  \vspace{-5mm}
  \end{center}
 \caption{
Results obtained for beam momentum of 3210.7~MeV/c corresponding to $Q$~=~0.76~MeV.
{\bf Top:} 
Missing mass spectrum from experimental data (dots), and simulations
(histogram). 
The simulated spectrum was normalized to the data.
{\bf Bottom:} 
Open points indicate the number of measured events of 
elastically scattered protons.
Solid points denote fit result of differential cross sections determined by
the EDDA Collaboration~\cite{edda2004}
with luminosity as the only free parameter.
}
 \label{lumi_fig}
\end{figure}

\begin{figure}[h!]
  \begin{center}
    \includegraphics[width=0.47\textwidth]{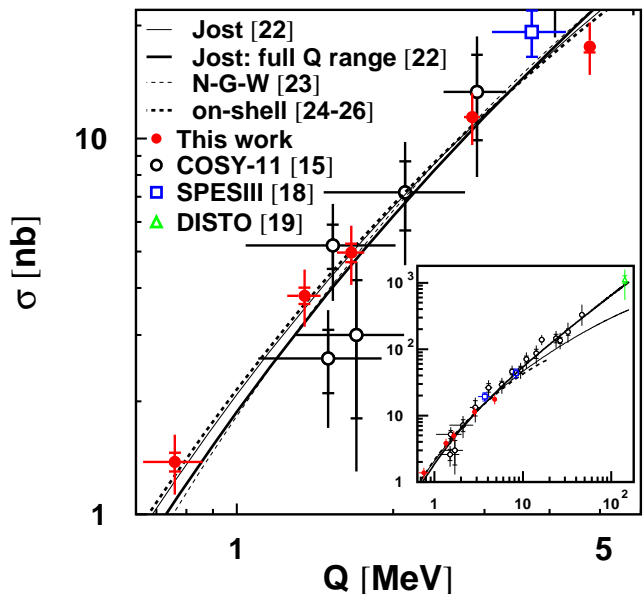}
  \end{center}
 \vspace{-7mm}
 \caption{
The total cross sections for the $pp\to pp\eta'$ reaction as a function of the excess energy.
Solid circles
represent new results reported in this article and results from previous experiments are shown with symbols as indicated in the legend. 
The statistical and systematic errors are separated by dashes.
The superimposed curves show results of fits with the $\eta'$-proton scattering length as a free parameter and parametrizing the $pp$-FSI
enhancement factor as in 
Refs.~\cite{noyes995,noyes465,naisse506} (thick dashed line), 
inverse of the squared Jost function~\cite{Druzhinin} (thin solid line)
and Niskanen-Goldberger-Watson model~\cite{Shyam} (thin dashed line). 
The thick dashed line is shown only in the range of applicability of 
the formula used for the enhancement factor~\cite{noyes995}.
For comparison the thick solid line shows result of the fit obtained for the whole $Q$ range with $pp$-FSI parametrization 
from Ref.~\cite{Druzhinin}.
The small plot shows the excitation function 
up to $Q=180$~MeV where the thin solid and thin dashed curves overlap.
 \label{excitation_function}
         }
 \vspace{-5mm}
\end{figure}

\begin{figure}[h!]
  \begin{center}
    \includegraphics[width=0.47\textwidth]{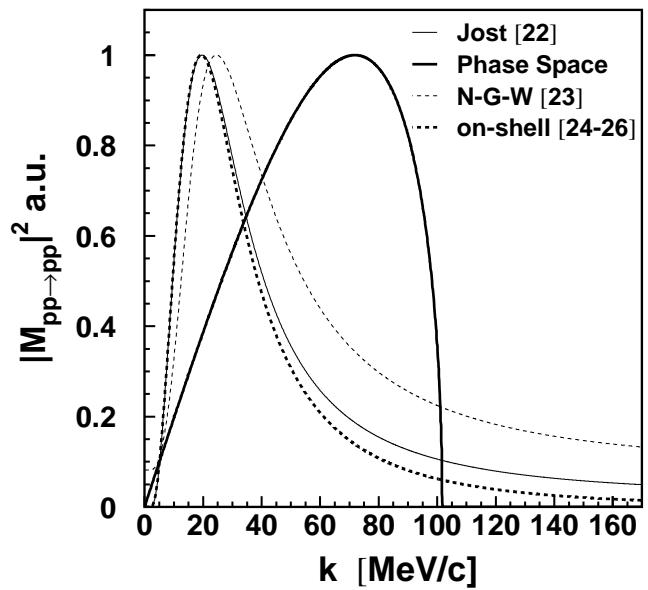}
  \end{center}
 \vspace{-7mm}
 \caption{Square of the proton-proton scattering amplitude calculated as a function of $k$, the proton 3-momentum in the proton-proton subsystem, 
parametrized as in Refs.~\cite{noyes995,noyes465,naisse506} 
(thick dashed curve), 
as inverse of the squared Jost function~\cite{Druzhinin}
(thin solid)
and using the Niskanen-Goldberger-Watson 
model~\cite{Shyam}
(thin dashed). 
The thick solid
curve shows the phase space $k$
distribution for $Q$ = 11 MeV. 
All the curves are arbitrarily normalized to unity at maximum. 
 \label{FSImodels}
         }
 \vspace{-5mm}
\end{figure}

In COSY--11 the $\eta'$ meson was produced in $p$-$p$ 
collisions of the COSY proton beam with an internal 
hydrogen cluster target.
The four-momenta of outgoing protons from the
$pp \to pp X$ 
reaction were measured in two drift chambers and
scintillator detectors 
and the $\eta'$ meson was identified via the
missing mass technique~\cite{Brauksiepe,eryk_prl}.
The low emittance proton beam combined with 
the high missing mass resolution of the COSY--11 detector
allowed measurements very close to the kinematic threshold
where the signal-to-background ratio 
increases due to the more rapid reduction of the phase space 
for multi-meson than for single meson production~\cite{eryk_prl}.
The measurement was conducted at five excess energies in the range
$Q$~=~0.76~MeV to $Q$~=~4.78~MeV.  
The determination of the absolute value of $Q$ was based on 
the position of the $\eta'$ signal in the missing mass spectra.
(A typical missing mass spectrum is shown in the top plot of 
 Figure~\ref{lumi_fig}).
$Q$ was determined with a precision of 0.10~MeV, 
where 0.06~MeV is due to the uncertainty of the $\eta'$ 
meson mass~\cite{PDG}
and 0.04~MeV comes from the possible misalignment 
of the relative setting of the detection system components 
and the center of the region of the beam and target overlap.
The latter was monitored by 
the measurement of elastically scattered protons~\cite{Moskal5}.
The experiment was designed to reduce the spread of 
excess energy to a negligible level
by the use of a rectangular collimator in the target setup 
so the width of the target stream was equal to 0.90~mm
while crossing the proton beam.
Due to the known dispersion of the COSY beam, this width is equivalent to an effective beam momentum spread of 
$\pm 0.06$~MeV/c corresponding to 0.02~MeV spread of 
excess energy $Q$.
The size of the target stream was monitored by a dedicated wire device with an accuracy of 0.05~mm~\cite{ErykPhD}
and in addition it was controlled 
independently by measuring elastically scattered protons.
The number of registered $pp\to pp$ events as a function of 
the protons scattering angle together with the known differential 
cross section for this process~\cite{edda2004} allows total luminosity determination as presented in the bottom plot of 
Figure~\ref{lumi_fig}.
The luminosity was determined for each beam momentum separately. 
The total luminosity for all measurements amounts to about 
2.55 pb$^{-1}$.
Due to the high statistics of the detected $pp\to pp$ events, 
the error in the determination of luminosity is less than 
0.05\% and can be neglected in the analysis below. 
Here we conservatively take the systematic uncertainty of 
the data from the EDDA Collaboration used for the normalization~\cite{edda2004}.
The number of identified $\eta'$ mesons was derived from 
fit of the 
simulated missing mass spectra to the experimental ones 
after background subtraction
(see the top plot of Figure~\ref{lumi_fig})
with the normalization as the only free parameter.
The background was determined experimentally~\cite{eryk_prl,ErykPhD}.
The geometrical acceptance and reconstruction efficiency 
for the $pp\to pp\eta'$ process 
were estimated based on simulations including experimentally determined properties of the COSY--11 detector
\cite{Brauksiepe}
and taking into account the final state interaction (FSI) of 
the outgoing protons~\cite{Moskal9}. 
Final results with statistical and systematic uncertainties 
are collected in Table~\ref{tab:result} and shown in 
Figure~\ref{excitation_function}.
Systematic uncertainties were estimated taking into account differences in the final result obtained by
(i) applying different methods of 
events counting, including variation of the range used 
for the background and signal counting and various binnings 
(3\%), 
(ii) different methods of background subtraction 
(7\%)~\cite{eryk_prl}, 
(iii) taking into account possible geometrical misalignment of 
the relative positions of the detection system components (2\%), 
(iv) uncertainty in the reconstruction efficiency of two close 
proton tracks (9\%)~\cite{Moskal8},
and (v) the relative uncertainty in the EDDA data sets used 
for the luminosity determination (2.5\%).
Because of the very small excess energies, 
the variation of the result due to different models of 
the proton-proton FSI (see Figure~\ref{FSImodels}) was found 
to be negligible.

\begin{table}[b!] 
\begin{center}
\begin{tabular}{ c|r }
$Q$~$\left[MeV\right]$ & $\sigma(pp\to pp\eta')$~$\left[nb\right]$ \\
\hline
0.76~$\pm$~0.10 &  1.38~$\pm$~0.08~$\pm$~0.17  \\
1.35~$\pm$~0.10 &  3.82~$\pm~$0.19~$\pm$~0.47  \\
1.66~$\pm$~0.10 &  4.97~$\pm$~0.28~$\pm$~0.61  \\
2.84~$\pm$~0.10 & 11.41~$\pm$~0.40~$\pm$~1.39  \\
4.78~$\pm$~0.10 & 17.58~$\pm$~0.64~$\pm$~2.15  
\end{tabular}
\end{center}
  \caption{Cross sections for the $pp\to pp\eta^{\prime}$ reaction at the five measured excess energies.
The excess energy $Q$ is tabulated  with the absolute systematic uncertainty and the cross section values are given with the statistical and systematic uncertainties, respectively.
}
  \label{tab:result}
\end{table}

Based on the data from previous experiments~\cite{eryk_acta,Moskal4,Moskal8,Khoukaz,PKlaja,Hibou,Balestra}
and the close to threshold total cross sections reported in Table~\ref{tab:result},
we have extracted the $\eta'$-proton scattering length. 
To this end the experimental excitation function 
for the $pp \to pp\eta'$ reaction was compared to 
the results of calculations taking into account 
proton-proton and $\eta' p$ interactions, 
where the real and imaginary parts of the $\eta' p$ 
scattering length were varied as free parameters.
At threshold the distance probed by the 
$pp\to pp\eta^{\prime}$ reaction is determined by the
momentum transfer between colliding nucleons and equal 
to about 0.2~fm,
whereas the typical range of the strong nucleon-nucleon or 
meson-nucleon interaction is of the order of a few Fermi. 
In addition the energy range considered in this article 
is two orders of 
magnitude smaller than the four-momentum transfer (1~GeV) 
governing the production amplitude. 
Therefore, the calculations were carried out using a Watson-Migdal approximation~\cite{watson} and the complete transition matrix
element of the $pp\to pp\eta'$ reaction was factorized as
\begin{equation}
\left|M_{pp\to pp\eta'}\right|^2 \approx \left| M_{0}\right|^2\cdot \left| M_{FSI}\right|^2~.
\end{equation}
Here $\left| M_{0}\right|^2$ 
represents the total short range production amplitude and 
$\left| M_{FSI}\right|^2$ denotes the final state interaction enhancement factor. 
Exact Fadeev calculations for the dynamics of three-body 
$pp\eta^{\prime}$ final states are 
unavailable.
Therefore, the enhancement factor 
for the $pp \eta'$ system was approximated
assuming the factorization of 
$M_{FSI}$ into two-particle scattering amplitudes~\cite{Moskal9},
\begin{equation}
M_{FSI} = M_{pp}(k_{1}) \times M_{p_{1}\eta'}(k_{2}) \times M_{p_{2}\eta'}(k_{3})~.
\label{row1}
\end{equation}
Here $k_{1}$ denotes the proton momentum in the proton-proton 
rest frame and $k_{2}$ and $k_{3}$ are the 3-momenta of 
$\eta^{\prime}$ and proton in the proton-$\eta'$ subsystems.
For the estimation of the proton-proton enhancement factor we have used the inverse of the squared Jost function~\cite{Druzhinin}.
To estimate the model dependence of the result
two other extreme solutions for 
the proton-proton enhancement factor were considered: 
the Niskanen-Goldberger-Watson parametrization~\cite{Shyam} 
and the square of the on-shell amplitude of the proton-proton scattering
calculated in the frame of the optical potential 
with phase shift 
including strong and Coulomb 
interactions~\cite{noyes995,noyes465,naisse506}.

The proton-proton and $\eta^{\prime}$-proton invariant mass
distributions determined for the $pp \to pp\eta^{\prime}$ 
reaction at an excess energy of $Q = 16.4$ MeV show an 
enhancement which may indicate a non negligible P-wave contribution from the proton-proton subsystem~\cite{PKlaja,kanzo}. 
Therefore in order to avoid a bias on the result from distortion of higher partial waves we restrict the extraction of the scattering length only to the range $Q$~$<$~11~MeV.
This limitation minimizes also the dependence of the result on the 
$pp$-FSI model and reduces the corresponding systematic uncertainty.
Moreover the low energy range used in the analysis allowed 
us to parametrize the $\eta' p$ FSI enhancement factor with 
the scattering length approximation
\begin{equation}
M_{\eta' p}=\frac{1}{1-ika_{\eta' p}}~
\label{F_petap}
\end{equation}
where $a_{\eta'p}$ is the scattering length of the $\eta' p$ interaction treated as a free parameter
in the analysis.\\
To determine $a_{\eta' p}$ we have constructed the following 
Neyman $\chi^{2}$ statistics
\begin{eqnarray}
\nonumber
\chi^2\left({\rm Re}(a_{p\eta'}), 
 {\rm Im}(a_{p\eta'}),\alpha\right) 
= \sum_{i=1}^{17}\frac{\left(\sigma_{i}^{expt}
- \alpha\sigma_{i}^{m}(a_{p\eta'})\right)^2} {\left(\Delta\sigma_{i}^{expt}\right)^2}
\\
\label{eqchi2_mh}
\end{eqnarray}
where $\sigma_{i}^{expt}$ denotes the $i$-th experimental total cross section measured with the statistical uncertainty 
$\Delta\sigma_{i}^{expt}$ and $\sigma_{i}^{m}$
stands for the calculated total cross section 
normalized with a factor $\alpha$
which is treated as an additional parameter of the fit. 
$\sigma_{i}^{m}(a_{p\eta'})$ was calculated for each excess energy 
$Q$
integrating Eq.(1) over the available phase space~\cite{pawelknot}. 
The best fit to the experimental data corresponds to
\begin{eqnarray}
\nonumber
\mathrm{Re}(a_{p\eta'}) &=&  0.00~\pm~0.43_{stat}~\mathrm{fm}~~~\mathrm{(syst.~error~negligible)}\\
\mathrm{Im}(a_{p\eta'}) &=& 0.37^{~+0.02_{stat}~+0.38_{sys}}_{~-0.11_{stat}~-0.05_{sys}}~\mathrm{fm}~. 
\label{chi2resB}
\end{eqnarray}
The statistical uncertainties in this case were determined at 
the 70\% confidence level taking into account that we have 
varied three parameters~\cite{james}. 
The systematic uncertainties due to the parametrization of the proton-proton interaction used in the analysis were estimated 
as the maximal difference between
the result obtained in Eq.~\ref{chi2resB}
and that determined using the two other $pp$-FSI models. 
For the real part of $a_{\eta' p}$ 
the differences obtained by applying different models are negligible.

It is interesting to compare these results with theoretical expectations and with recent studies based on the $\eta'$-nucleus optical potential.
In the Quark Meson Coupling model (QMC) \cite{etaA}
one calculates the in-medium meson masses and 
corresponding effective in-medium meson-nucleon scattering 
lengths through coupling the light quarks in the meson
to the scalar isoscalar $\sigma$ 
(and also $\omega$ and $\rho$) mean fields in the nucleus. 
For 20 degrees $\eta$-$\eta'$ mixing angle, 
QMC predicts the $\eta'$ mass shift to be -37 MeV at 
nuclear matter density $\rho_0$, 
corresponding to the real part of the effective 
$\eta'$-nucleon scattering length being 0.5 fm. 
This mass shift is very similar to the recent determination of 
the $\eta'$-nucleus optical potential by the CBELSA/TAPS
collaboration from studies of 
$\eta'$ photoproduction from Carbon \cite{nanova1}.
The $\eta'$-nucleus optical potential 
$V_{\rm opt} = V_{\rm real} + iW$
deduced from these photoproduction experiments is 
$
V_{\rm real} (\rho_0)
= -37 \pm 10 (stat.) \pm 10 (syst.) \ {\rm MeV}
$
which is equal to the meson mass shift in medium
and
$
W(\rho_0) = -10 \pm 2.5 \ {\rm MeV}
$.
Larger mass shifts, 
downwards by up to 80-150 MeV, were found in NJL 
\cite{hirenzaki}
and linear sigma model calculations \cite{jido1}. 
Each of these
theoretical models prefers a positive sign for the real 
part of $a_{\eta' N}$ in medium. 
A chiral coupled channels calculation performed with
possible scattering lengths with real part between 0 and 1.5 fm
is reported in \cite{osetetaprime}.
A free-space scattering length close to zero was found in a 
coupled channels fit to $\eta'$ scattering processes \cite{nakayama}.
The energy and density dependence of the $\eta'$- 
(and also $\eta$-) nucleon scattering lengths is a open
topic of investigation \cite{gal1}.
If one assumes no density and energy dependence of the $\eta'$
nucleon scattering length, then the value obtained in Eq.(5) 
is consistent with the QMC result \cite{etaA} and 
disfavors the expectations in~\cite{hirenzaki,jido1}.

In summary, the close to threshold excitation function
for the $pp\to pp\eta^{\prime}$ reaction was determined down to an excess
energy of $Q=0.76$~MeV
with the precision $\Delta{Q}=0.10$~MeV 
improved by more than a factor of five with respect to previous measurements.
The achieved resolution enabled the first quantitative extraction 
of the scattering length for the $\eta^{\prime}$ proton interaction in free space. 
Most importantly, the extracted value of the real part of the scattering length is found to be independent of the proton-proton 
FSI model
in the close to threshold energy range (up to 11 MeV) used in the fit.
\begin{acknowledgments}
We acknowledge support
by the Polish National Science Center through grants No. 2011/03/B/ST2/01847, 2011/01/B/ST2/00431, 
by the FFE grants of the Research Center J\"{u}lich,
by the Austrian Science Fund (FWF) through grant P23753,
the European Commision through European Community-Research Infrastructure Activity under FP6 project
Hadron Physics (contract number RII3-CT-2004-506078),
and by the Polish Ministry of Science and Higher Education through grant No. 393/E-338/STYP/8/2013.
\end{acknowledgments}
\vspace{-0.5cm}

\end{document}